\documentstyle[12pt,a4,epsf]{article}
\begin{document} 
\bibliographystyle{plain} 
\input{epsf} 
{\Large \bf \noindent Statistical Mechanics of a
Two-Dimensional System with Long Range Interactions} 
\vskip 2 truecm
\noindent \footnote{CNRS-Laboratoire de Physique Th\'eorique
de l'Ecole Normale Sup\'erieure, 24 rue Lhomond, F-75231, Paris Cedex 05,
France.}David S. Dean and \footnote{ Dipartmento di Fisica and INFN, Universit\`a di Roma, La Sapienza, 
P. A. Moro 2, 00185 Roma, Italy.}Giorgio Parisi
\pagestyle{empty} 
\vskip 1 truecm \noindent{\bf Abstract:} We analyse the statistical physics
 of a two dimensional lattice based gas with long range interactions. 
The particles interact in a way analogous to Queens on a chess board. 
The long range nature of the interaction gives the mathematics of the problem a simple geometric structure which simplifies  both the analytic and numerical 
study of the system. We present some analytic calculations for the statics 
of the problem and also we perform Monte Carlo simulations which exhibit a
dynamical transition between a high temperature liquid regime and a 
low temperature glassy regime exhibiting aging in the two time correlation 
functions.

\vskip 1 truecm
\noindent  October 1997
\vskip 1 truecm
\noindent  LPTENS 97/48
\newpage
\pagenumbering{arabic} 
\pagestyle{plain} 
\def\half{{1\over 2}}
\def\OO{\Omega}
\def\sech{{\rm sech}}
\def\n{{\newline}}
 \def\aa{\alpha}
 \def\bk{{\bf k}}
 \def\bkp{{\bf k'}}
 \def\bqp{{\bf q'}}
 \def\bq {{\bf q}}
 \def\EE{\Bbb E}
 \def\EEx{\Bbb E^x}
 \def\EEo{\Bbb E^0}
 \def\LL{\Lambda}
 \def\PP{\Bbb P^o}
 \def\rr{\rho}
 \def\SS{\Sigma}
 \def\ss{\sigma}
 \def\ll{\lambda}
 \def\dd{\delta}
 \def\ww{\omega}
 \def\ll{\lambda}
 \def\DD{\Delta}
 \def\DDt{\tilde {\Delta}}
 \def\kr{\kappa\lb \LL\rb}
 \def\PPx{\Bbb P^{x}}
 \def\gg{\gamma}
 \def\kk{\kappa}
 \def\tt{\theta}
 \def\bs{\hbox{{\bf s}}}
 \def\bh{\hbox{{\bf h}}}
 \def\lb{\left(}
 \def\rb{\right)}
 \def\prt{\tilde p}
\def\pt{\tilde {\phi}}
 \def\bb{\beta}
 \def\hal{{1\over 2}\nabla ^2}
 \def\bg{{\bf g}}
 \def\bx{{\bf x}}
 \def\bu{{\bf u}}
 \def\by{{\bf y}}
 \def\hag{{1\over 2}\nabla}
 \def\beq{\begin{equation}}
 \def\eeq{\end{equation}}
 \def\cosech{\hbox{cosech}}
\section{Introduction}
It is believed that one of the essential ingredients for the formation
of a glassy phase is the presence of frustration along with a large 
number of metastable states. The glass transition is a dynamical one,
if one considers ordinary glass one sees the glass phase because
on rapid cooling the system does not manage to find the crystalline
structure of silicon dioxide. Instead the system becomes trapped on experimental
time scales in a state of higher energy than the crystalline state. This 
phase is characterised by very slow dynamics and often exhibits aging
phenomena in many observable quantities such as the correlation and response
functions. A good introduction to the problem of the glass transition
may be found in \cite{got}. Much of the understanding of glassy or out
of equilibrium dynamics has been made via analogy with the dynamics of
spin glasses and other systems with quenched disorder (for a review
see \cite{bocukume}). Ideas from the theory of spin glasses have been
tentatively put forward to understand the glass transition \cite{mepa}.
In this paper we put forward a new model which we hope will be of use
as a test of these ideas on systems without quenched disorder. We 
will emphasise its attractiveness as a model from both analytic and
simulational points of view.

A well known puzzle to chess players is the Eight Queens Problem. 
On a standard chess board it may be summarised as the problem of 
arranging eight queens such that no queen is in a position to take
another queen at the next move. The interested reader will find that 
it is a non trivial task to find such an arrangement. 
In this paper we shall analyse a  gas model based on the Eight Queens Problem.

The attraction of this model is that the long range nature of the
interactions facilitates a rather straight forward and efficient
implementation Monte Carlo dynamics for computer simulations. In
addition we shall see that, whilst we cannot solve the statics
exactly, various approximation schemes become analytically much simpler.

The space of the model is an $L\times L$ lattice with periodic boundary 
conditions imposed to facilitate easier analytic and numerical analysis, the
original problem of course does not have periodic boundary conditions.
On this lattice are $N$ particles, the energy of a given configuration
is given by the Hamiltonian

\beq
 H = \sum_{i\neq j}^N \delta^L_{x_i,x_j} + \delta^L_{y_i,y_j} + 
\delta^L_{x_i - y_i,x_j -y_j} + \delta^L_{x_i + y_i, x_j + y_j} ,
\label{eq:H}
\eeq

where the position of the $i$th particle is denoted by the pair $(x_i,y_i)$.
The superscript $L$ on the $\delta$ indicates that they are the standard
Kronecker delta functions but with arithmetic modulo $L$. One can
see that for a periodic chess board the solution of the Eight Queens Problem
is given the zero energy configurations of the Hamiltonian system above, the
first two terms in the Hamiltonian represent the row and column constraints
and the second two represent the constraint on the left to right and right
to left diagonals respectively.

In certain cases it is rather easy to find the zero energy states of the 
system. Let us consider the case where $L$ is prime. As an ansatz we shall
take the configuration 
\beq x_i = i \hbox{ and } y_i = pi, \eeq
for each $i$ between $1$ and $N$ and where $p$ is another a number  different
from $L$ and $1$. Clearly the column constraint is satisfied automatically.
 To violate the row constraint we would have to have
\beq p(i-j)\equiv L \eeq
where $\equiv$ will always be taken to indicate equality modulo $L$, the fact
that $L$ is prime means that this may only be satisfied when $i\equiv j$. 
Violation of the first diagonal constraint would mean that
\beq (p-1)(i-j) \equiv L \eeq
This would imply that $i\equiv j$ or $p-1 \equiv L$. Violation of the second 
diagonal constraint would imply
\beq (p+1)(i-j) \equiv L, \eeq
 hence either $i\equiv j$ or $p+1 \equiv L$. 
Therefore when one can find $p$ such that $p+1$ and $p-1$ are not equal
to zero modulo $L$ then the above construction does indeed give a zero
energy state. This analysis therefore demonstrates the existence of zero
energy states for all $L$ prime strictly greater than $3$.

This analysis is useful for our study of the dynamics; the ground states
above are our analogy of the crystalline state of silicon dioxide
when comparison is made with real glass. 

\section{Mean Field Analysis}
One may carry out a geometrical mean field Boltzmann type of analysis
to determine the energy per particle of the system. Consider the
configuration about one particle at the point $O$ which is
fixed. Ignoring the effects of correlations the particle at $O$ has
two types of sites surrounding it, those with which it interacts directly,
i.e. those which are on the same row, column or diagonal which we
shall call type $A$, and those with which it does not interact
directly which we shall call type $B$. Denote by $N_{X\to Y}$ the
number of sites of type $Y$ that  interact with a given
particle of on a site of type $X$ ($X,Y \in \{ A,B, O\}$). 
Simple counting yields
\begin{eqnarray}
N_{O\to A}&=& 1 \hbox{, by definition} \nonumber \\
N_{A\to A}&=&  N+7 \hbox{, interaction with $N-2$ on the line
interacting with $O$} \nonumber\\  &{\rm and}& \hbox{intersection 
with 3 $A$ s on the 3 other lines} \nonumber \\
N_{B \to A}&=& 3N - 12 \hbox{, using $N_{O\to A}+N_{A\to A}+N_{B \to A} = 4N
- 4$}   \nonumber \\
N_{O\to B}&=& 0 \hbox{, by definition} \nonumber \\
N_{A\to B}&=& 12 \hbox{, 4 lines with 3 intersections each} \nonumber \\
N_{B\to B}&=& 4N - 16 \hbox{, using $N_{O\to A}+N_{A\to A}+N_{B \to A} = 4N
- 4$}  
\end{eqnarray}
The Boltzmann equations for the average number of particles in sites of
type  type $A$ and $B$ are
\beq 
\langle A \rangle = \lambda \exp\left( - (N+7)\beta\langle A \rangle -
(3N- 12)\beta  \langle B \rangle  -\beta\right) 
\eeq 
and
\beq
\langle B \rangle = \lambda \exp\left( -12 \beta\langle A \rangle -
(4N - 16)\beta \langle B \rangle \right) .
\eeq
Here $\lambda$ is a Lagrange multiplier factor enforcing the over all
particle number to be $N$ i.e so that

\beq 
1 + 4(N-1) \langle A\rangle + (N^2 - 4N +3)\langle B\rangle = N,
\eeq
which simplifies to
\beq 
4\langle A\rangle + (N-3)\langle B\rangle = 1.
\eeq
In the limit $N \to \infty$ one finds
\beq \langle B\rangle = 1/N \hbox{ and } \langle A\rangle = a/N ,\eeq
where $a$ is finite.
Consequently
\begin{eqnarray}
\langle B \rangle &=& \lambda( 1 - {a\over N}(1-e^{-\beta}))^{12}\times
(1- {1\over N}(1 - e^{-\beta}))^{4N - 16} \nonumber \\
&\to& \lambda \exp\lb -4(1 - e^{-\beta}) \rb \nonumber \\
&=& {1\over N}.
\end{eqnarray}
Hence
\beq 
\lambda = {1\over N} \exp\lb -4(1 - e^{-\beta}) \rb.
\eeq
Taking similarly the large $N$ limit for the equation for $\langle
A\rangle$ yields
\beq
a = \exp\lb (1-e^{-\beta})(1-a) -\beta \rb,
\eeq
and the energy $E$ per particle is given by 
\beq
E(\beta) = 2 a(\beta).
\eeq
For infinite temperature one finds that $E(0) = 2$ and for low
temperatures $E(\beta) \sim 2 e^{1-\beta}$. Hence the mean field
calculation does not contradict the fact that there are zero energy
ground states for $N$ prime.

\section{The Hypernetted Chain Approximation}
Perhaps one of the most successful approximations in gas and liquid
theory is the hypernetted chain (HNC) approximation \cite{hamc}, which is an
integral equation which resums a certain class of diagrams in the
virial expansion. Normally this integral equation must be resolved
numerically and its resolution is rather difficult. However we shall
see here that for our model the resulting equation is drastically
simplified in the large $N$ limit.

The matrix form of the HNC equation is
\beq \log(1 + h_{ij}) = -\bb V_{ij} + \rho \sum_{k} h_{ik}(h_{kj} -
\log(1 + h_{kj}) -\bb V_{kj}) \eeq
where 
\beq h_{ij} = {1\over\rho^2}(\langle \rho_i \rho_j \rangle - \rho
\delta_{ij} -\rho^2).
\eeq
Here $\rho_i$ is the density at the site $i$ and $\rho = \langle
\rho\rangle$ in homogeneous systems. In the present model $\rho =
1/N$. 
One may write the interaction matrix as
\beq V_{ij} = 3\delta_{ij} + W_{ij} \eeq
The important simplification comes from noting that (using the
counting from the previous section)
\beq W^2 = (3N -9)I + (N-6)W + 12U ,\eeq
where $I$ is the identity matrix and $U$ is the matrix with each
element equal to $1$. Also
\beq U^2 = N^2 U \hbox{ and } UW = WU = (4N - 3)U .\eeq
Hence the elements $W, I, U,$ form a closed algebra. In general we may
therefore represent an element in this algebra by a triple
\beq A \equiv (a_1,a_2,a_3) = a_3 U + (a_2 - a_3)W + (a_1 -a_2)I \eeq
where 
\begin{eqnarray}
A_{ij} &=&  a_1 \hbox{ when } i=j \nonumber \\ 
A_{ij} &=&  a_2 \hbox{ when } i\neq j \hbox{ and } V_{ij}=1 \nonumber \\ 
A_{ij} &=&  a_3 \hbox{ when } V_{ij} = 0.
\end{eqnarray} 
In this notation the product of two elements $A = (a_1,a_2,a_3)$ and
$B = (b_1,b_2,b_3)$ is given by $C = AB = (c_1,c_2,c_3)$ where
\begin{eqnarray}
c_1 &=& a_3 b_3 N^2 + (a_1-a_2)(b_1-b_2) + (a_2-a_3)(b_2-b_3)(4N -3) 
\nonumber \\ &+& (a_3(b_2-b_3) +b_3(a_2-a_3))(4N -3)+
(a_3(b_1-b_2)+b_3(a_1-a_2))\nonumber \\ &+& ((a_1-a_2)(b_2-b_3) + (b_1-b_2)(a_2-a_3))
\end{eqnarray} 

\begin{eqnarray}
c_2 &=& a_3 b_3 N^2  + (a_2-a_3)(b_2-b_3)(N + 6) \nonumber \\ &+&
 (a_3(b_2-b_3) +b_3(a_2-a_3))(4N-3)+ (a_3(b_1-b_2)+b_3(a_1-a_2))
\nonumber \\
 &+& ((a_1-a_2)(b_2-b_3) + (b_1-b_2)(a_2-a_3))
\end{eqnarray}
and
\begin{eqnarray}
c_3 &=& a_3 b_3 N^2  + 12 (a_2-a_3)(b_2-b_3) \nonumber \\ &+& (a_3(b_2-b_3)
+b_3(a_2-a_3))(4N -3)+ (a_3(b_1-b_2)+b_3(a_1-a_2)) 
\end{eqnarray}
Using this algebra and taking the limit $N\to \infty$ at the end of
the calculation one finds that in this representation the HNC reduces to 
\beq \log(1 + h_1) = -4\beta + {4h_2^2 + h_3 -7h_2 h_3 + 3 h_3^2\over 
 1 + h_2-h_3}, \eeq
\beq \log(1 + h_2) = -\beta  +{h_2^2 -h_2 h_3 + h_3\over 1 + h_2-h_3}, \eeq
and
\beq \log(1+h_3) = h_3. \eeq
The final equation is trivial giving $h_3 = 0$ and hence
\beq h_1 =
\exp(-4\bb + {4h_2^2\over 1 + h_2})-1 
\eeq
with
\beq 
\log(1 + h_2) = -\beta  +{h_2^2\over 1 + h_2}.
\eeq
For $\bb$ small one finds $h_2 \sim -\bb$ and for
$\bb$ large $h_2 \sim -1 + 1/(\beta + 2-\log(\beta) +
O(\log(\beta)/\beta))$.
The energy per particle is then given by
\beq
 E = {N\over 2}\sum_{i = 1}^{N^2}\langle \rho_0 \rho_i\rangle V_{0i} =
2(h_2 + 1).
\eeq
Hence, at high temperatures, $E \sim 2(1-\beta)$ as in the mean field 
calculation, but $E
\sim 2/(\beta + 2 -\log(\beta) +O(\log(\beta)/\beta))$ at low temperatures.

\section{Monte Carlo Simulation}

The dynamics for the Hamiltonian (\ref{eq:H}) were simulated using a
Monte Carlo method using the Metropolis algorithm a with sequential
update on the particles. Two variants of the dynamics have been analysed, 
firstly a non local dynamics where a particle may move to any site on the
board and secondly a local random walk dynamics where a particle may move
to any of its eight nearest neighbours on the lattice. One of the advantages
of the current model is that the energy may be expressed as a function of the 
occupation numbers of each row ($N_r(k)$), column ($N_c(k)$), left to right 
diagonal ($N_{d-}(k)$) and right to left diagonal ($N_{d+}(k)$). This speeds up greatly the calculation of the energy change at each move. In this notation
 it is easy to see that 
\beq E = {1\over 2 N}\sum_{k = 1}^N ( N_r(k)^2 + N_c(k)^2 +N_{d-}(k)^2 
+N_{d+}(k)) - 2 \label{eq:EN}
\eeq
This leads to a reduction by a factor of $N$ of the time needed to compute
the energy change with respect to a calculation using the particle 
positions. The form of the energy given by equation (\ref{eq:EN})
demonstrates that if one were to neglect the diagonal interactions one
would arrive at a model similar to two independent Backgammon
models \cite{rit}, but at negative temperature. The Hamiltonian is not strictly
the same but the tendency is to put all the particles in different
boxes rather than to put them into single box which is the ground
state of the Backgammon model. Here the free energy barrier clearly
have an energetic component as opposed to the Backgammon model where
their origin is entropic. We shall also see later that, in common with
the Backgammon model, the dynamical transition temperature appears to
be at $T=0$.

The system was started from a random initial configuration,
thus simulating a rapid quench from high temperature to the simulation
temperature.  In both the cases of non local and local dynamics the
energy as a function of time was measured.

In the case of non local dynamics  the particle-particle 
correlation function treating the particles as distinguishable. It is
defined by
\beq C(t,t') = {1\over N}\sum_i \delta_{x_i(t),x_i(t')}
\delta_{y_i(t),y_i(t')},
\label{eq:cp}
\eeq
note that $C$ is normalised to be one when $t=t'$
(note in keeping with convention $t'$ will always denote the earlier
time).  In the case of local dynamics it is interesting to measure the
correlation function which gives us information about the effective
particle diffusivities, this correlation function is normally denoted
by $B(t,t')$ in the literature and is given by
\beq 
B(t't') = {1\over N}\sum_i (x_i(t) - x_i(t'))^2.
\eeq
We shall discuss the results of the two dynamic types
separately.

\noindent{\bf Non Local Dynamics}

With non local dynamics one expects, at an intuitive level, that the
dynamics should be faster and less prone to be glassy with respect to
local dynamics. If one imagines a self consistent picture of a non
ordered phase, then a single particle moves in a potential generated
by the other particles. With non local moves trapping mechanisms at
the spatial level are less important.

The simulations we carried out were for systems of 50000 particles and
times of up to 60000 Monte Carlo sweeps through the entire
system. In addition we verified that for systems of size 48611 (which
is prime) the results were not significantly altered. 

In figure (1) is shown the results of the energy per 
particle measured
by the simulation for $t= 30000$. Up to values of $\beta = 5$ we found
that by $t= 30000$ the energy density had reached a well defined
asymptotic value. We found for values of $\beta \geq 6$ that at
$t=30000$ the system had not reached an asymptotic values of the
energy density and that the value of the energy density continued to
decay though very slowly. The values of $E$ shown on figure (1) for
$\beta \geq 6$ are the values measured at $t=30000$ and are not
equilibrium values. One sees from figure (1) that between $\beta = 5$
and $\beta = 6$ the measured energy flattens off rather suddenly in a
fashion reminiscent of glassy systems.

For vales of $\beta < 5$ and a minimum waiting time of 10000 we found
equilibrium behaviour for the correlation function. In this region
\beq C(t,t') = C(t-t') \approx A\exp({t-t'\over \tau_0}).\label{eq:Etau}\eeq
One may regard $\tau_0$ as a characteristic time scale, if this time
scale is greater than the observation time then one should expect out
of equilibrium behaviour. Shown in figure (2) is a plot of 
$\log(\tau_0(\beta))$ as measured by fitting the simulation data with
the formula (\ref{eq:Etau}) above. As one can see the plot is linear
and the best fit is $\tau_0(\beta) = 0.317\cdot \exp (1.796\beta)$, 
when $\beta$
is such that $\tau_0(\beta)$ is much less than the age of the system
one expects that equilibrium should be achieved. For an age of the
system of order 10000 one finds that the value of the temperature
where the age and $\tau_0(\beta)$ have the same age is $\beta \approx 5.7$
which is entirely consistent with the plateau observed in the energy
and with the breakdown of time translational invariance in the
correlation function.

For values of $\beta \geq 6$ we found that the function $C(t,t')$
was no longer time translational invariant, neither did it have an
exponential form. For values of $\beta \geq 7$ we found that to a reasonable
degree of accuracy the system exhibits perfect aging, i.e. $C(t,t')
= f(t/t')$, see for example the curves in figure (3) for $\beta = 10$. At $T=0$
one sees that the system also ages and the five curves shown collapse onto 
the same master curve when plotted against $t\over t_w$, this is shown in 
figure (4). This is presumably because at $T=0$ the characteristic
time $\tau_0$ becomes infinite and there is no interrupted
aging \cite{France1}. 
The scaling $t/t'$ is been observed in a variety of mean field
models \cite{paps1} and also in the phenomenological trap model
\cite{paps2}. The improvement of the perfect aging
scaling as one approaches  zero temperature was also observed by 
Ritort in Monte Carlo simulations of the Backgammon model \cite{rit}.

\noindent{\bf Local Dynamics}

Here we consider the more physical situation of local random walk dynamics.

Shown in figure (5) is the energy measured for the local dynamics for
a system of size 50000 particles up till a time of 2000 Monte Carlo
sweeps. Up till $\beta = 4$ the energy reaches an asymptotic value,
where as for $\beta > 4$ the energy continues to decrease. The values
shown in these cases is the minimum value of the energy attained. As
one sees clearly in the figure, this measured value is not monotonic
and actually increases after $\beta = 6.5$. For longer
simulation times this minimum in the measured dynamic energy 
shifts towards larger values of $\beta$. 

For values of the waiting time $t' = 10000,\ 20000, \ 30000, \ 40000$ up
till $\beta = 4$ one sees that
\beq B(t,t') = \kappa(\beta)(t-t') \eeq
and hence we are in an equilibrium  phase for these
temperatures and at this scale of age. 
The values of $\log(\kappa(\beta))$ are shown in figure (6) and one
sees that the curve is linear and one finds that $ \kappa(\beta) = 
0.34\cdot\exp(-1.08\beta)$. 

In the nonequilibrium  regime we find that one may fit the curves as
an anomalous diffusion but with a prefactor that depends on the
waiting time. The behaviour of the particles is subdiffusive and we
have fitted it with the form
\beq 
B(t,t') = c(\beta) {(t-t')^\alpha \over t'^\gamma},
\eeq 
where $\gamma = 2(1-\alpha)$.
This simple linear relation between the exponents for $(t-t')$ and $t'$
works rather well but we have no theoretical reason to expect this
result. An example of the rescaled curves for $\beta = 6$ is shown in 
figure (6). The values of $\alpha$ as a function of $\beta$ are shown 
in figure (7). To go to lower values of the temperature than those
shown is rather difficult as the noise in the measured values of
$B(t,t')$ becomes rather large, presumably because one is approaching
the zero temperature transition.

\section{Conclusions}

We have presented a preliminary study of a two dimensional particle
system which exhibits a dynamical transition between a gas/liquid and a
glassy phase. As expected from the dynamical nature of the transition
the temperature at which one sees this transition is dependent on both
the time scale of the observations and on the nature of the dynamics. 
The long range nature of the interactions leads to an compact
application of various standard approximation schemes for the statics -
particularly the hypernetted chain approximation which works
reasonably well in the high temperature phase. The model is also
attractive from the simulation point of view as the form of the
interactions leads to an efficient implication of the Monte Carlo
updating procedure.

\noindent{\bf Acknowledgments} We would like to thank Lorenza and Leonardo
Parisi for drawing our attention to the Eight Queens Problem.

\pagestyle{empty}
\vskip 2 truecm
\noindent{\bf List of Figure Captions}

\noindent{\bf Fig 1.} Nonlocal dynamics: Dynamic Energies for $L= 50000$ at $t = 30000$. Calculated static energies using the HNC and the mean field
approximation are also show.

\noindent{\bf Fig 2.} Nonlocal dynamics: $\log\tau_0(\beta)$ $L= 50000$
(with linear fit shown).

  \noindent{\bf Fig 3.} Nonlocal dynamics: $C(t,t_w)$ plotted against $t/t_w$ at $\beta =10$ for
$t_w = 5000,\ 10000, \ 15000, \ 20000, \ 25000$ ($L$ = 50000).

\noindent{\bf Fig 4.} Nonlocal dynamics: $C(t,t_w)$ plotted against $t/t_w$ at $T = 0$ for
$t_w = 20000,\ 40000, \ 60000, \ 80000, \ 100000$ ($L$ = 50000).

\noindent{\bf Fig 5.} Local dynamics: Dynamic Energies for $L= 50000$
at $t = 2000$.

\noindent{\bf Fig 6.} Local dynamics: $-\log(\kappa(\beta)$ in the
equilibrium regime (with linear fit shown).

\noindent{\bf Fig 7.} Local dynamics: Rescaled $B(t+t_w,t_w)$ for
$t_w = 10000,\ 20000, \ 30000, \ 40000$ at $\beta = 6$ with $\alpha$
taken to be $0.862$.

\noindent{\bf Fig 8.} Local dynamics: $\alpha(\beta)$ in the
nonequilibrium regime.

\newpage
\begin{figure}[htb]
\begin{center}\leavevmode
\epsfxsize=10 truecm\epsfbox{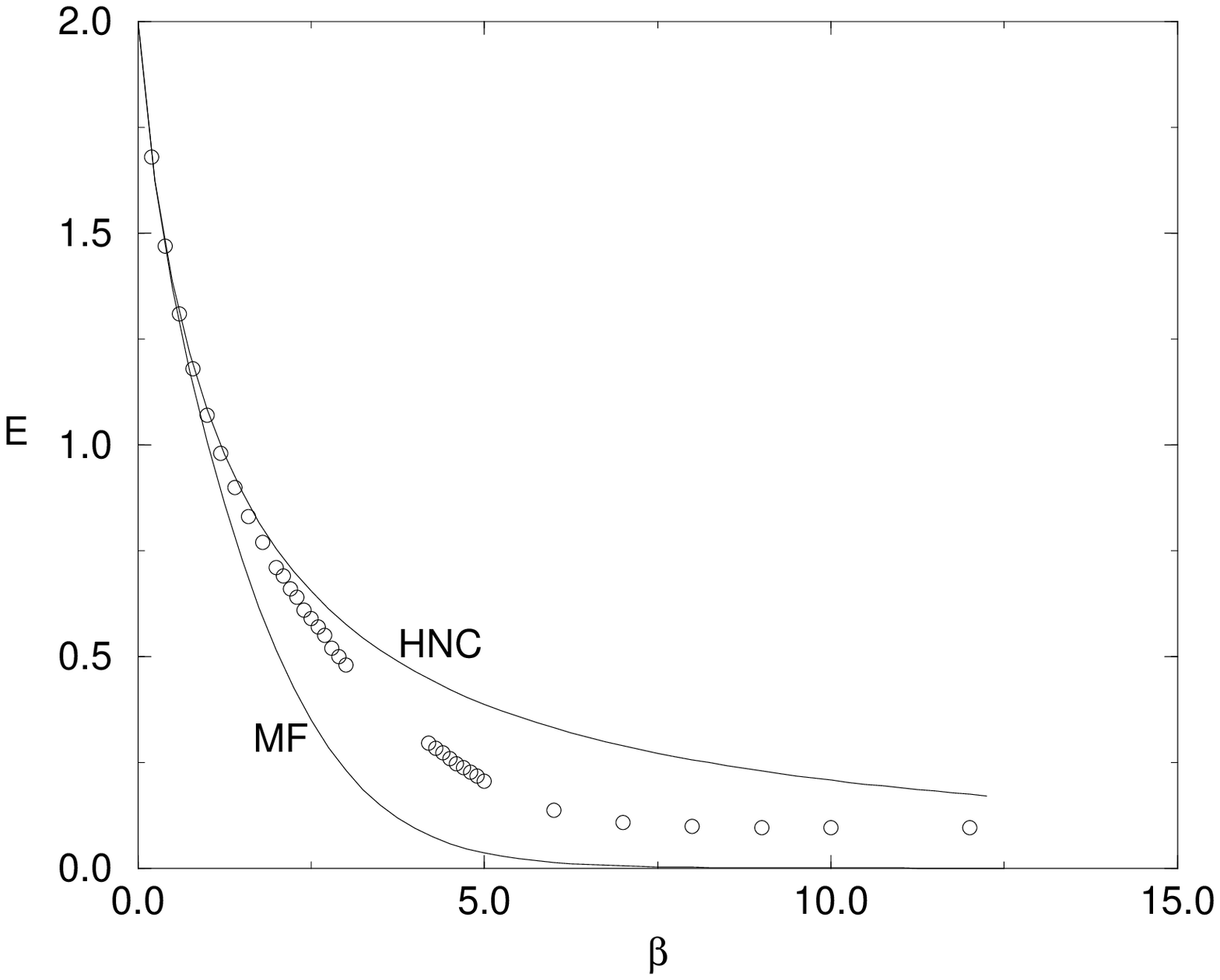}
\end{center}
\caption[]{}
\label{figure:Fig1}
\end{figure}

\begin{figure}[htb]
\begin{center}\leavevmode
\epsfxsize=10 truecm\epsfbox{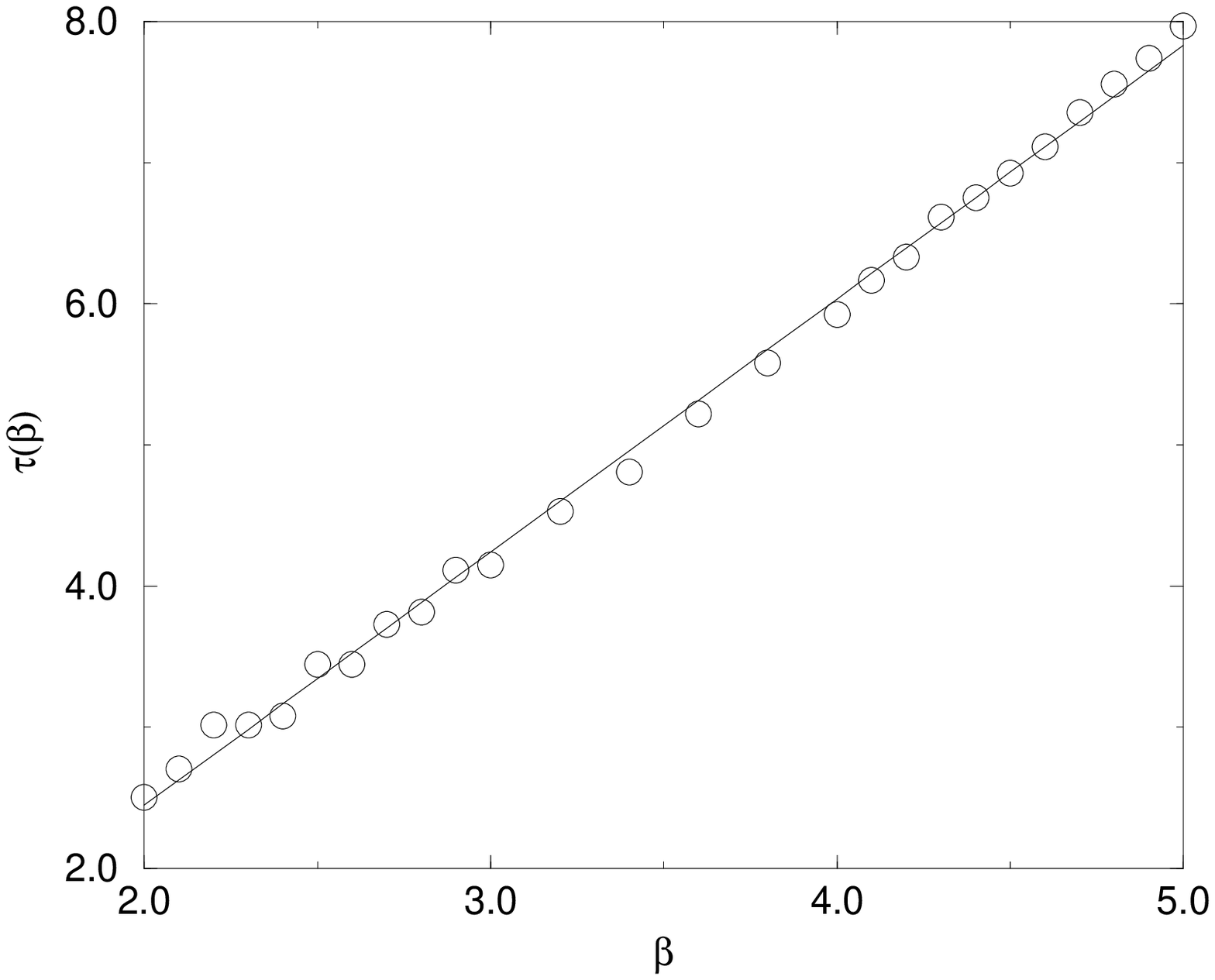}
\end{center}
\caption[]{}
\label{figure:Fig2}
\end{figure}

\begin{figure}[htb]
\begin{center}\leavevmode
\epsfxsize=10 truecm\epsfbox{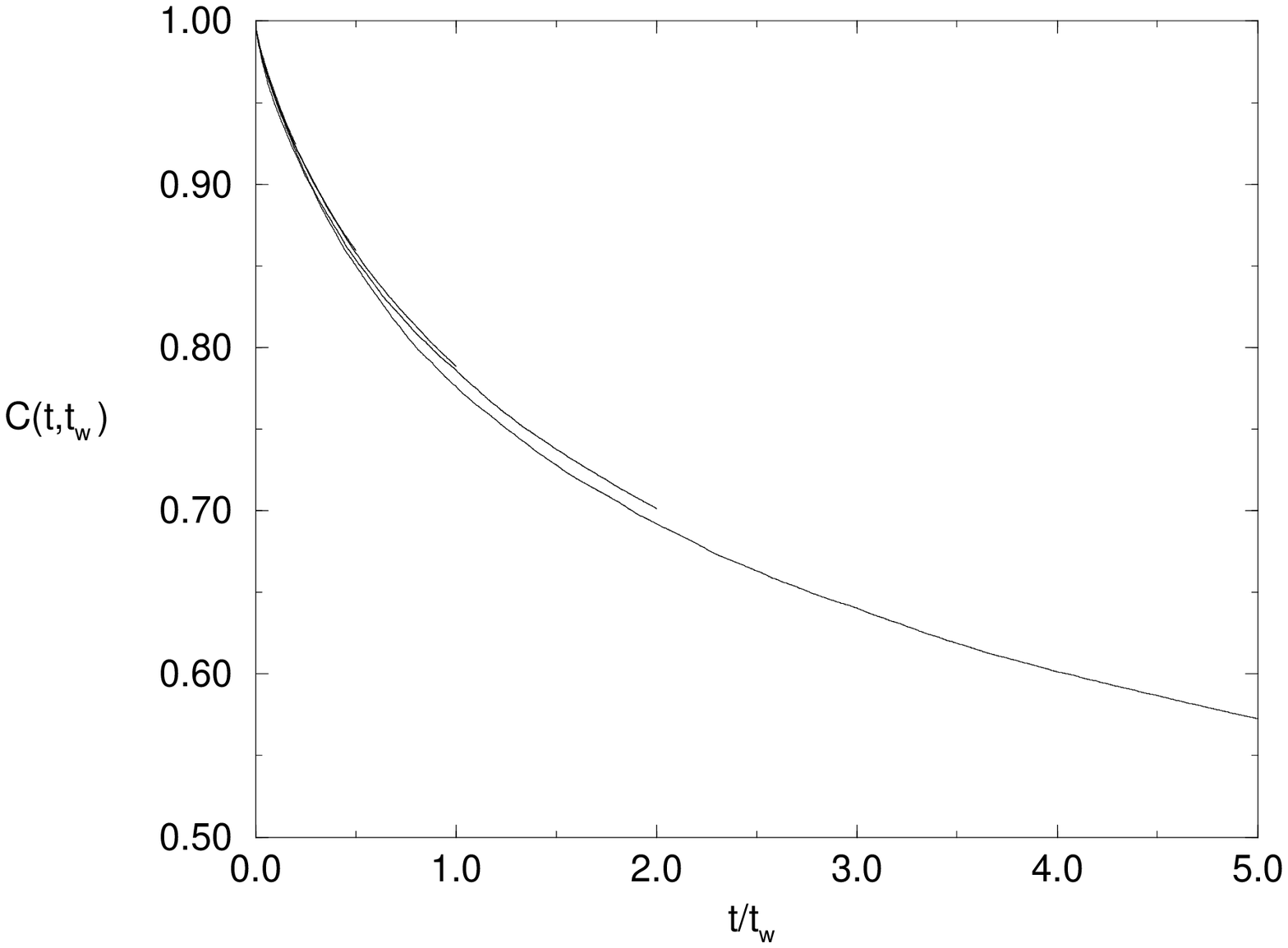}
\end{center}
\caption[]{}
\label{figure:Fig3}
\end{figure}

\begin{figure}[htb]
\begin{center}\leavevmode
\epsfxsize=10 truecm\epsfbox{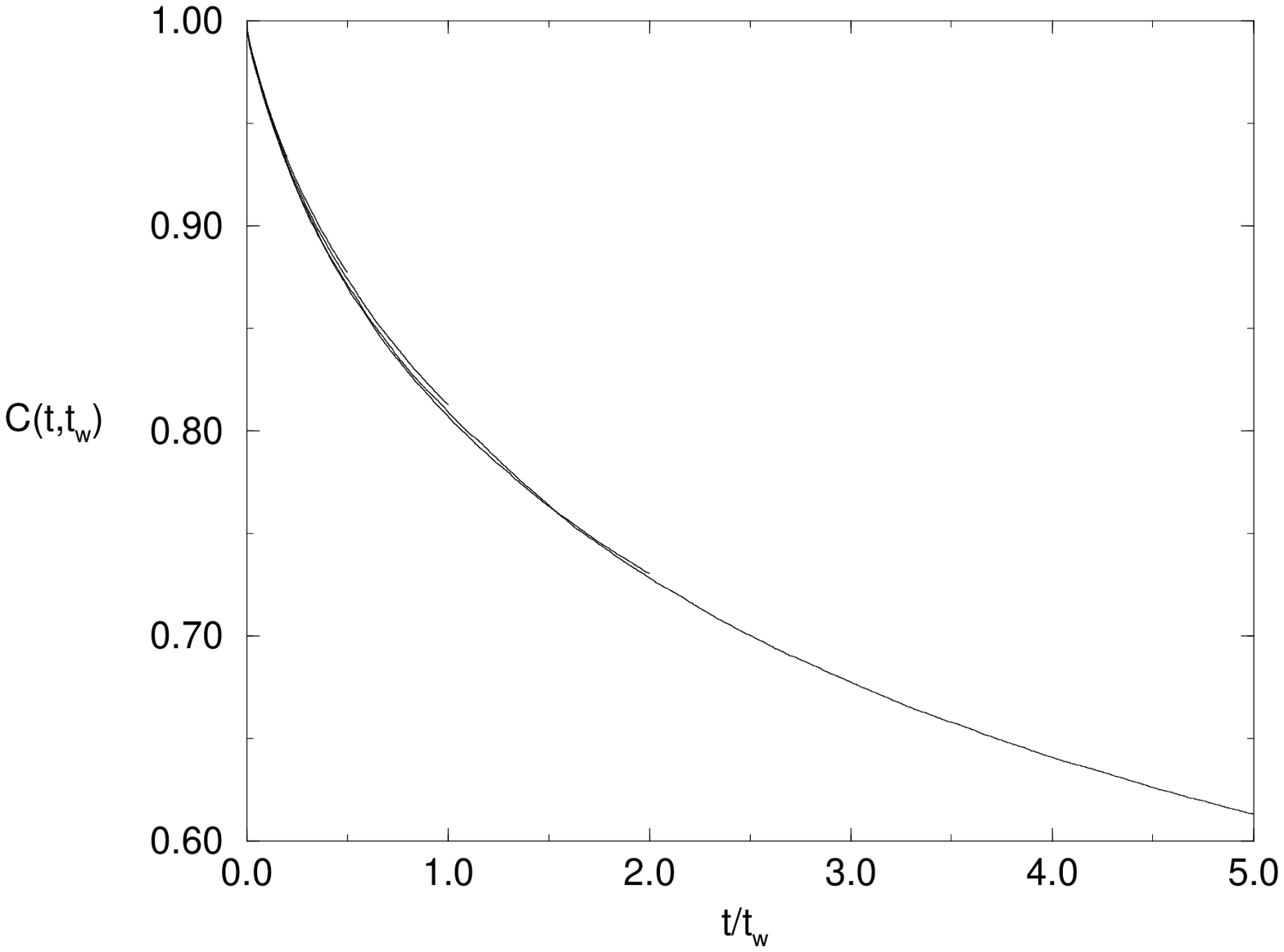}
\end{center}
\caption[]{}
\label{figure:Fig4}
\end{figure}

\begin{figure}[htb]
\begin{center}\leavevmode
\epsfxsize=10 truecm\epsfbox{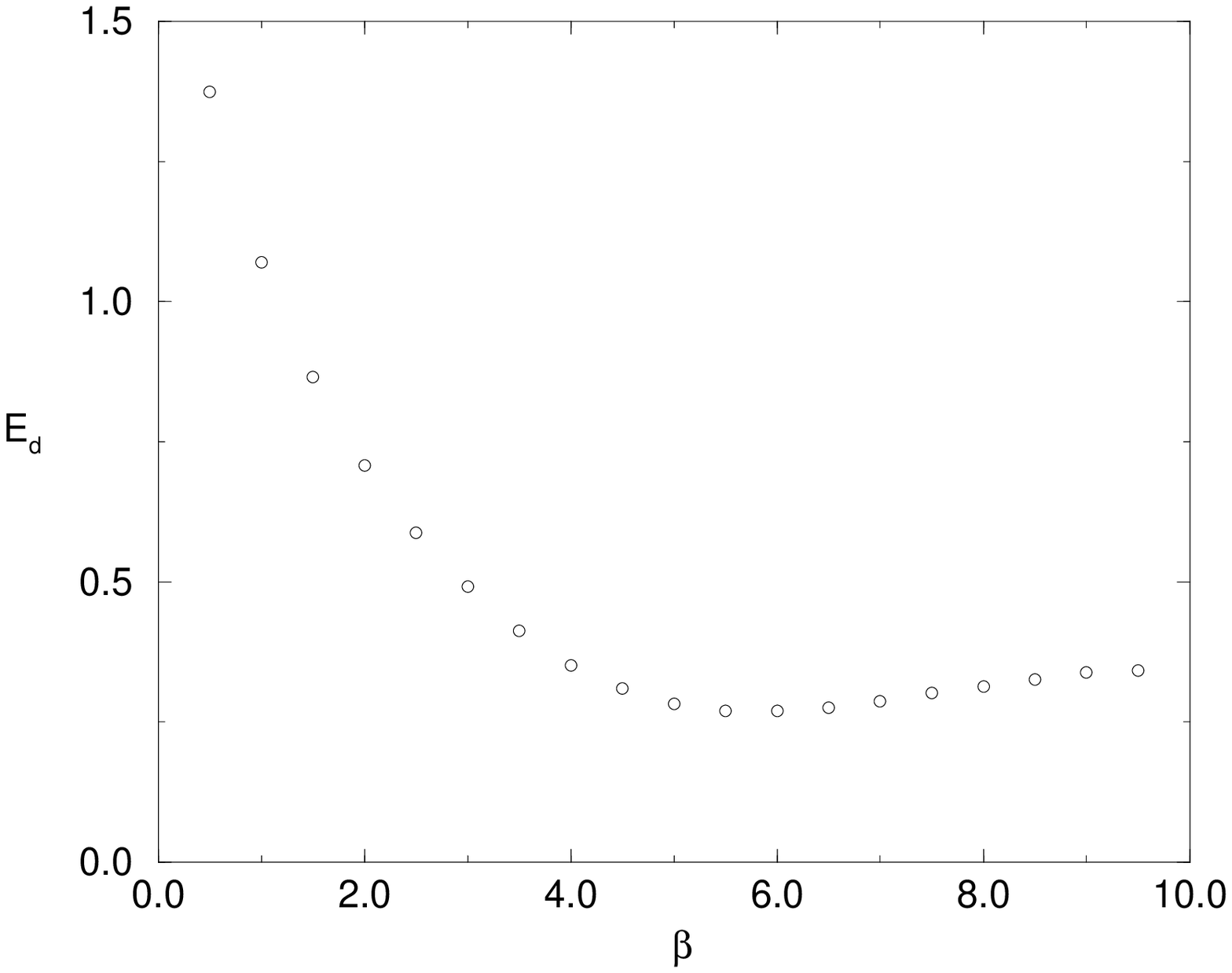}
\end{center}
\caption[]{}
\label{figure:Fig5}
\end{figure}

\begin{figure}[htb]
\begin{center}\leavevmode
\epsfxsize=10 truecm\epsfbox{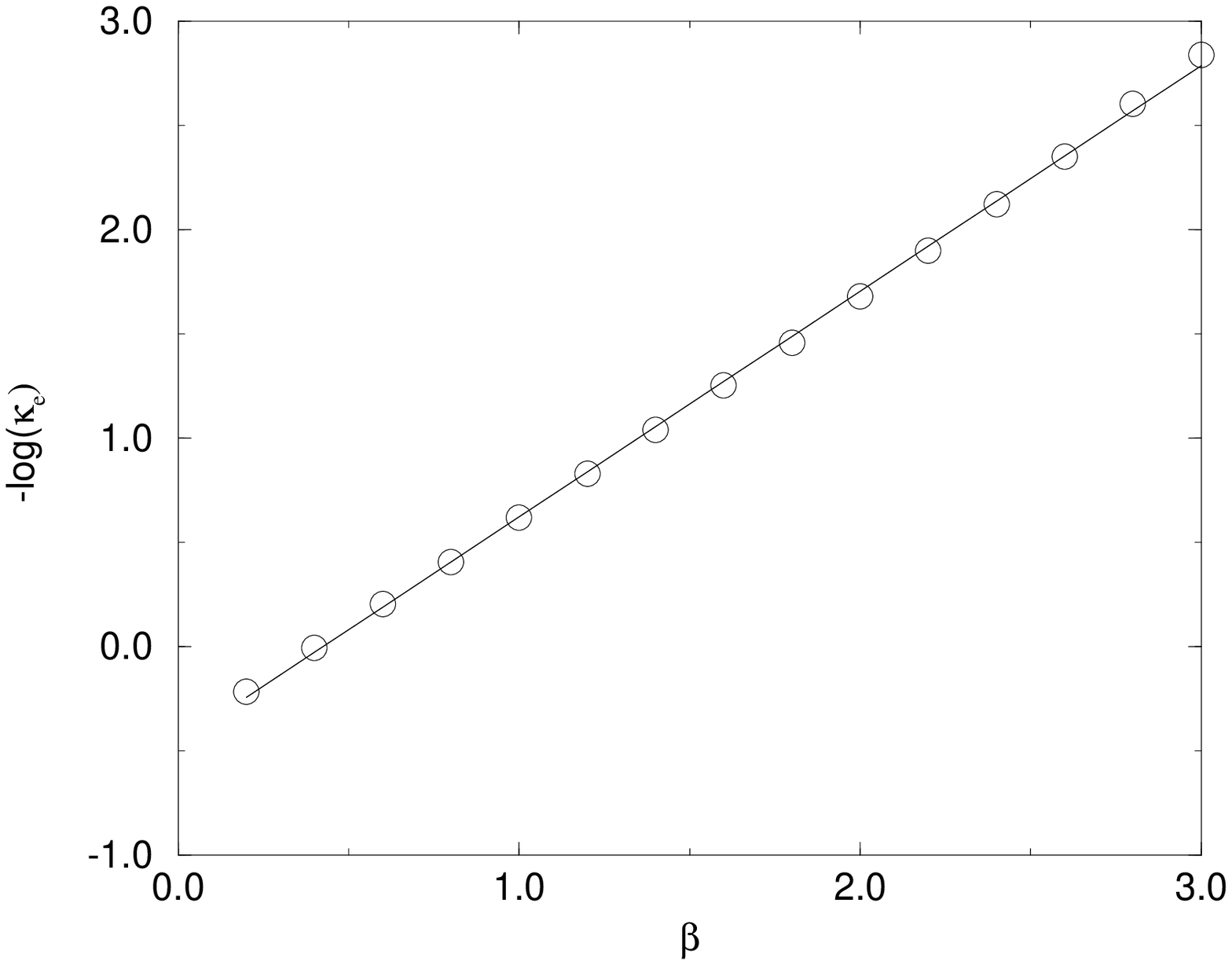}
\end{center}
\caption[]{}
\label{figure:Fig6}
\end{figure}

\begin{figure}[htb]
\begin{center}\leavevmode
\epsfxsize=10 truecm\epsfbox{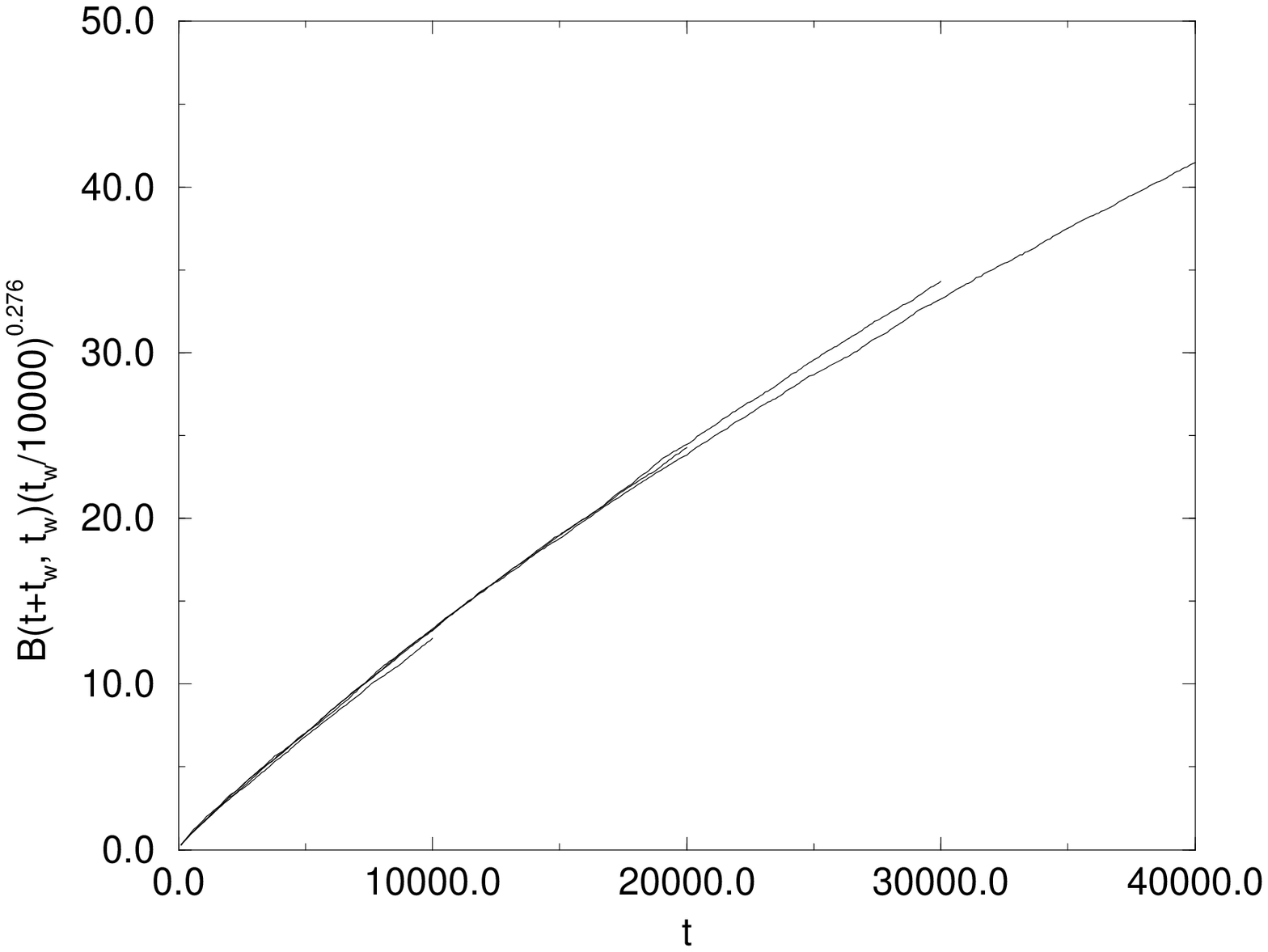}
\end{center}
\caption[]{}
\label{figure:Fig7}
\end{figure}
 
\begin{figure}[htb]
\begin{center}\leavevmode
\epsfxsize=10 truecm\epsfbox{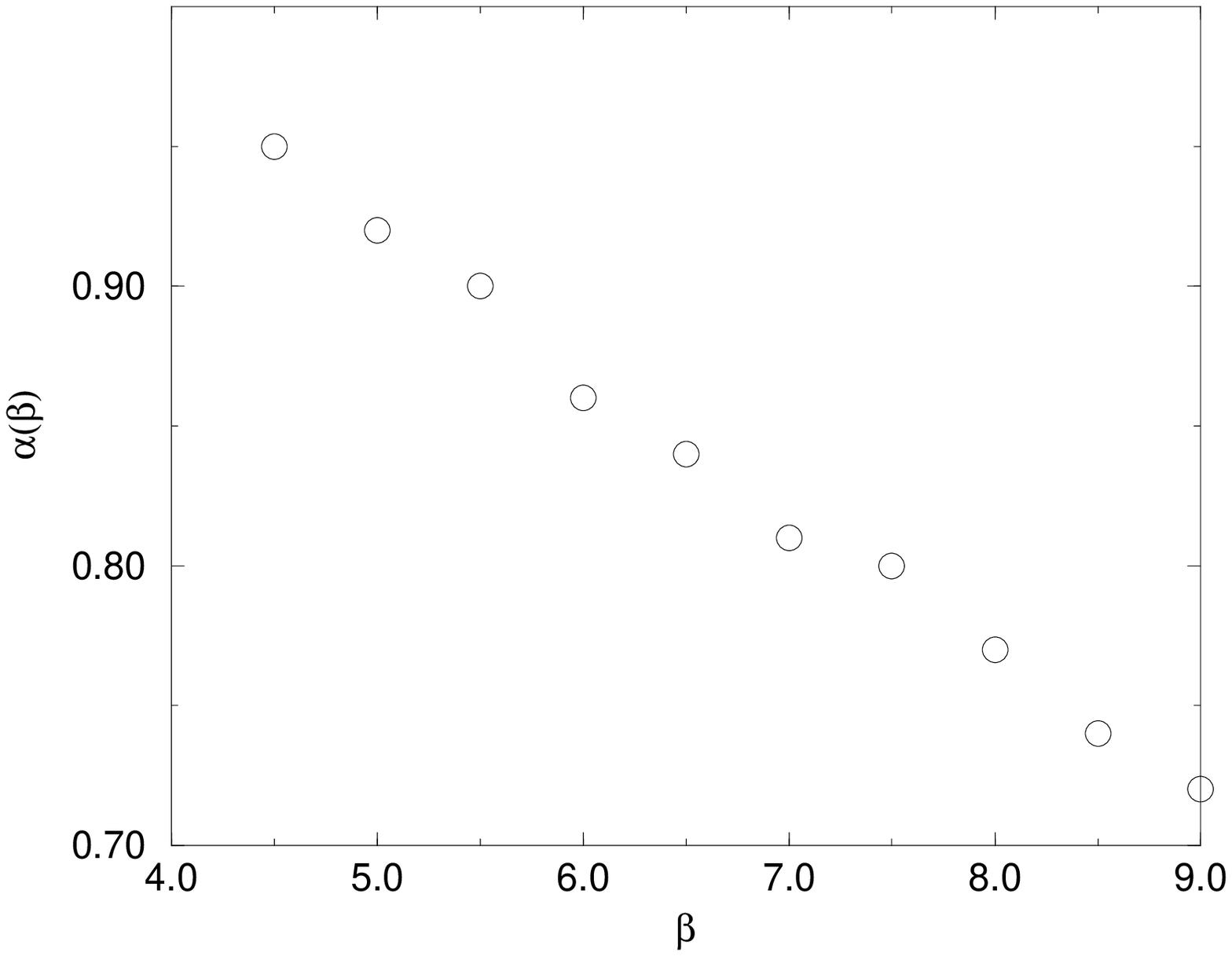}
\end{center}
\caption[]{}
\label{figure:Fig8}
\end{figure}

\end{document}